\documentclass[twocolumn,a4paper]{article}
\usepackage[colorlinks, urlcolor=blue,
	pdffitwindow, bookmarksopen,
	filecolor=blue,
	bookmarksopenlevel=0,
	pdfpagelayout=SinglePage, dvipdfm]{hyperref}
\usepackage{ncsp24} 
\usepackage{color}
\usepackage[dvipdfmx]{graphicx}
\usepackage{hyperref}

\usepackage{cite}
\usepackage{amsmath,amssymb,amsfonts}
\usepackage{algorithmic}
\usepackage{textcomp}
\usepackage{xcolor}
\usepackage{multirow}
\usepackage{dblfloatfix}

\begin{document}

\title{A Random Ensemble of Encrypted models for Enhancing Robustness \\ against Adversarial Examples}

\name{
 \begin{tabular}{c}
Ryota Iijima$^1$, Sayaka Shiota$^1$ and Hitoshi Kiya$^1$
 \end{tabular}
 }

\address{
\begin{tabular}{c}
$^1$Tokyo Metropolitan University\\
6-6 Asahigaoka, Hino-shi, Tokyo 191-0065, Japan\\
E-mail: iijima-ryota@ed.tmu.ac.jp, sayaka@tmu.ac.jp, kiya@tmu.ac.jp
\end{tabular}
}

\maketitle

\section*{Abstract}
Deep neural networks (DNNs) are well known to be vulnerable to adversarial examples (AEs).
In addition, AEs have adversarial transferability, which means
AEs generated for a source model can fool another black-box model (target model) with a non-trivial probability.
In previous studies, it was confirmed that the vision transformer (ViT) is more robust against the property of adversarial transferability than convolutional neural network (CNN) models such as ConvMixer, and moreover encrypted ViT is more robust than ViT without any encryption.
In this article, we propose a random ensemble of encrypted ViT models to achieve much more robust models.
In experiments, the proposed scheme is verified to be more robust against not only black-box attacks but also white-box ones than convention methods.

\vspace{-3mm}

\section{Introduction}
Deep neural networks (DNNs) have achieved great success in various computer vision tasks, and they have been deployed in many applications including security-critical ones such as face recognition and object detection for autonomous driving \cite{hitoshi2022an}.
However, DNNs are known to be vulnerable to adversarial examples (AEs), which fool DNNs by adding small perturbations to input images without any effect on human perception.
In addition, AEs designed for a source model can deceive other (target) models. This property, called transferability, make it easy to mislead various DNN models.
This is an urgent issue that has a negative impact on the reliability of applications using DNNs. \par
In previous studies, various methods were proposed to build robust models against AEs.
Adversarial training \cite{goodfellow2015explaining,kurakin2017adversarial, Hongyang2019theoratically} is widely known as a defense way against AEs, where AEs are used as a part of training data to improve the robustness against AEs.
However, it degrades the performance of models when test data is clean. \par
Another approach for constructing robust models is to train models by using images encrypted with secret keys \cite{maung2020encryption, aprilpyone2021block}.
The encrypted models were verified to be robust against white-box attacks if the keys are not known to adversaries.
Furthermore, the approach is effective in avoiding the influence of transferability \cite{tanaka2022on}.
However, this approach is still vulnerable to black-box attacks because black-box ones do not need to know any secret key. \par
Accordingly, we propose a random ensemble of encrypted ViT models that has been inspired by model encryption \cite{maung2020encryption,aprilpyone2021block, tanaka2022on} and an ensemble of models \cite{pang2019improving, yang2020dverge}.
An ensemble of encrypted models was discussed as one of adversarial defenses \cite{aprilpyone2021block,maungmaung2021ensemble}, but the model is not robust yet against black-box attacks when the ensemble model is used as the source model.
The proposed method with a random ensemble allows us not only to use robust models against black-box attacks but
to also utilize robust ones against white-box attacks.
In an experiment, the effectiveness of the proposed method is verified on the CIFA10 dataset by using a benchmark attack, called AutoAttack \cite{croce2020reliable}

\vspace{-3mm}

\section{Related Work}
\subsection{Adversarial Examples}
Depending on the ability of adversaries, there are three types of attacks: white-box attacks \cite{goodfellow2015explaining,madry2018towards,carlini2017towards,moosavi2016deepfool}, black-box attacks \cite{andriushchenko2020square}, and gray-box attack. 
Adversaries have complete knowledge of the target model and data information in white-box settings.
In black-box settings, adversaries have no knowledge of the target model other than the input images and the outputs of the model.
In black-box settings, AEs are often generated without any information on the target model, in which AEs are designed by using a property of AEs called adversarial transferability.
Between white-box and black-box settings, there are gray-box attacks that mean that the adversary knows something about the model.
Furthermore, AEs can be categorized into two types in terms of the goal of adversaries.
Target attacks mislead the output of models to a specific class.
In contrast, non-targeted attacks aim to mislead models to an incorrect class. \par
AutoAttack \cite{croce2020reliable} was also proposed to evaluate the robustness of defense methods against AEs in an equitable manner.
AutoAttack consists of four parameter-free attack methods: Auto-PGD-cross entropy (APGD-ce), Auto-PGD-targe (APGD-t), FAB-target (FAB-t) \cite{croce2020minimally}, and Square attack \cite{andriushchenko2020square} (see Table \ref{tab:autoattack}).
In RobustBench  \cite{croce2021robustbench}, many defense methods have been evaluated on the basis of AutoAttack.
In this paper, the proposed method is evaluated by using AutoAttack.

\begin{table}[t]
    \centering
    \caption{Attack Methods in AutoAttack}
    \vspace{3mm}
    \begin{tabular}{c|c|c}
         \multirow{2}{*}{Attack}    & White-box (W)/    & Targeted (T)/     \\
                                    & Black-box (B)     & Non-targeted (N)  \\
         \hline
         APGD-ce                    & W                 & N                 \\
         APGD-t                     & W                 & T                 \\
         FAB-t                      & W                 & T                 \\
         Square                     & B                 & N
    \end{tabular}
    \label{tab:autoattack}
\end{table}

\subsection{Key-based Defense}
Maung et al. proposed block-wise transformation with a secret key to build robust models against AEs \cite{aprilpyone2021block}.
In the method, a DNN models is trained by using images encrypted with a secret key.
The process of training a model is shown as follows.

\begin{enumerate}
    \item Divide each training image into blocks with a size of $M \times M$.
    \item Encrypt each block by using a transformation algorithm with a secret key.
    \item Integrate encrypted blocks to generate encrypted images.
    \item Train a model by using the encrypted images to obtain an encrypted model.
\end{enumerate}
To get an estimation result from the encrypted model, a test image has to be encrypted with the same secret key. \par

Furthermore, Tanaka et al. pointed out that some block-wise transformation methods including pixel shuffling can reduce the transferability among encrypted models if the keys are different, even when the models have the same architecture \cite{tanaka2022on}.
In this paper, we use pixel shuffling proposed in \cite{aprilpyone2021block} for block-wise image encryption.

\vspace{-3mm}

\section{Proposed Method}
\subsection{Overview}
\begin{figure}[t]
\begin{center}
  \includegraphics[width=70mm]{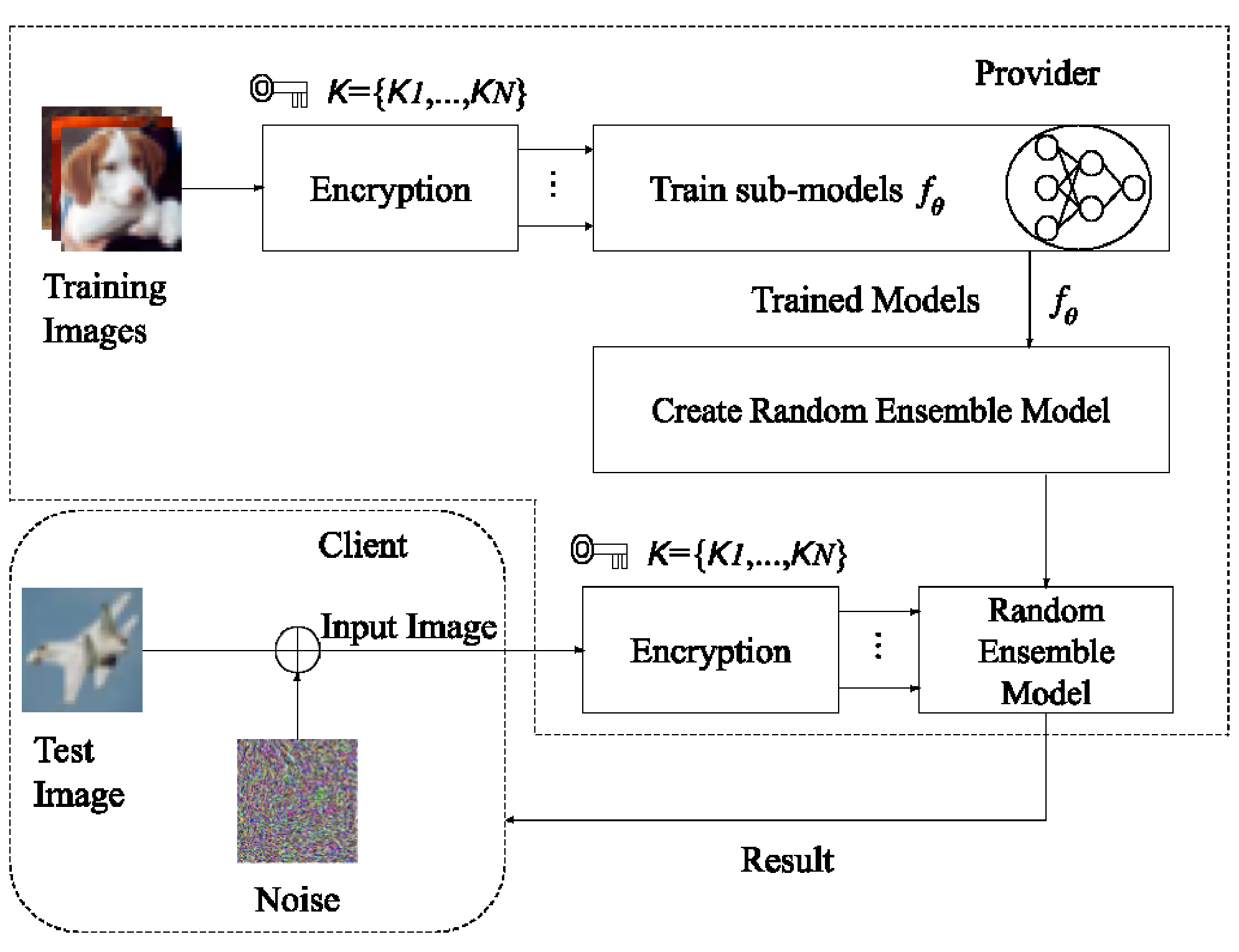}
  \caption{Framework of  proposed scheme}
 \label{fig:framework}
\end{center}
\end{figure}

Figure \ref{fig:framework} shows the framework of the proposed scheme.
At first, a provider trains $N (\geq 4)$ sub-models with images encrypted with $N$ secret keys $K=\{K_{1},..., K_{N} \}$, respectively, where pixel shuffling \cite{aprilpyone2021block} is used for encryption in this paper, and a different secret key is assigned to each sub-model for the encryption.
Next, the provider builds a random ensemble by using the sub-models.
In testing, the provider encrypts every test image with keys $K$ to generate $N$ encrypted test images from each test image, and $N$ encrypted images are input to the random ensemble model to get an estimation result.


\subsection{Random Ensemble Model}
\begin{figure}[t]
\begin{center}
  \includegraphics[width=70mm]{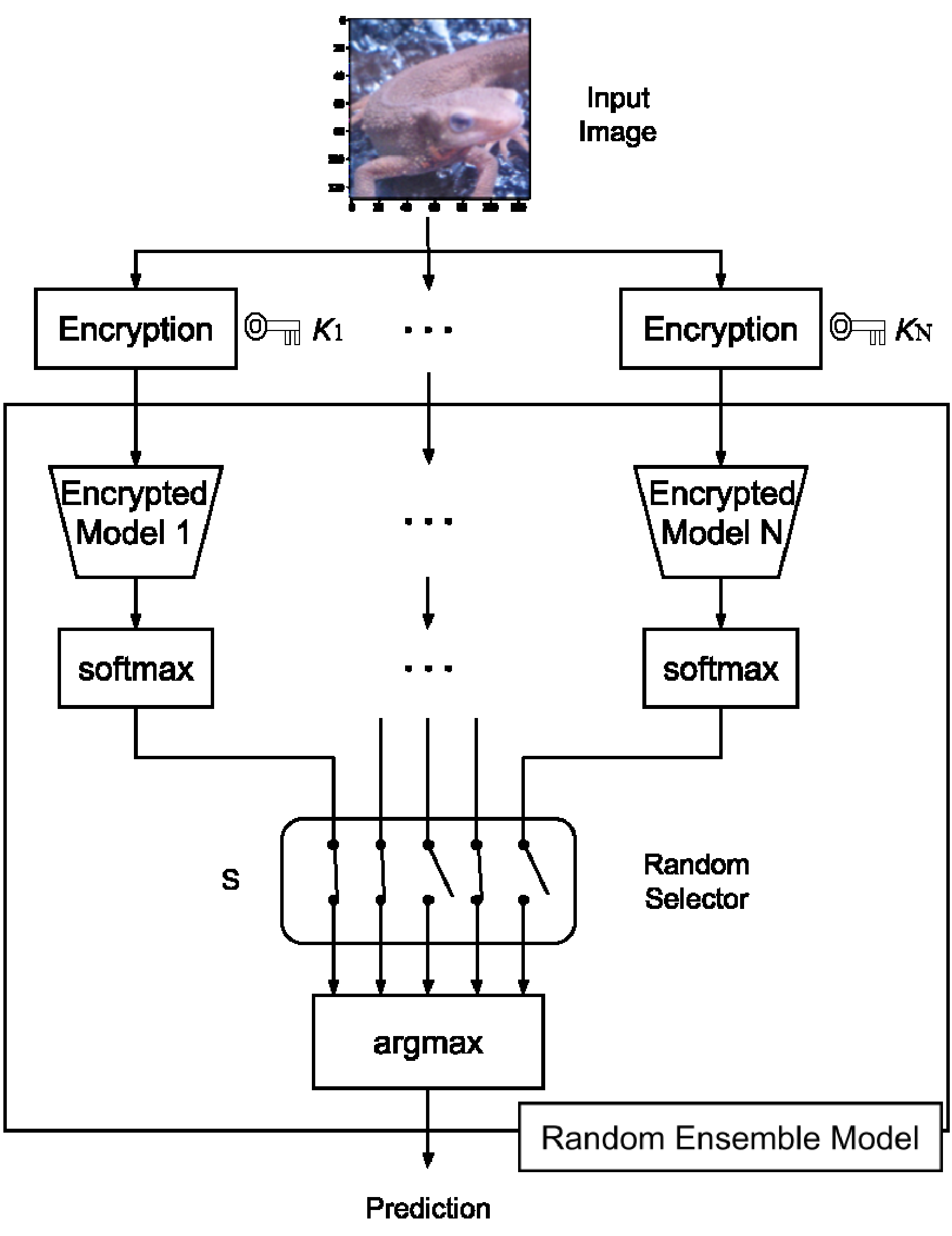}
  \caption{Overview of proposed method}
 \label{fig:randomensemble}
\end{center}
\end{figure}

Figure \ref{fig:randomensemble} shows the details of a random ensemble of $N$ encrypted sub-models where ViT is chosen as the architecture of sub-models due to the low transferability between encrypted sub-models \cite{mahmood2021robustness, pang2019improving, yang2020dverge, tanaka2022on}.
Below is steps for getting a prediction result from a test image.
\begin{enumerate}
    \item Generate $N$ encrypted images from a test image by using $N$ keys used for the encrypted sub-models.
    \item Input every image encrypted with each key to the sub-model encrypted with the same key.
    \item Select $S$ results $( 3 \leq S \leq N )$ from the sub-models randomly. 
    \item Determine a final output by the average of $S$ results.
\end{enumerate}
Even if the same image is input to the model, the method will provide a different estimation value.
Accordingly, it is expected that the feature makes various attacks including black-box ones more difficult.


\vspace{-3mm}

\section{Experiment}
The random ensemble was demonstrated to be effective in enhanced robustness against AEs on the CIFAR-10 dataset \cite{krizhevsky2009learning}.

\subsection{Experimental Setup}
The CIFAR-10 dataset consists of $60,000$ images with $10$ classes ($6,000$ images for each class) where $50,000$ images are for fine-tuning and $10,000$ images are for testing.
All images were resized to $224 \times 224 \times 3$ to fit the input to ViT, and they were scaled to $[0,1]$ as a range of the values.
We used pre-trained ViTs with a patch size of $P=16$ as sub-models where the pre-training was carried out with ImageNet-21k \cite{dosovitskiy2021an}.
We fine-tuned ViTs for $5,000$ epochs with the stochastic gradient descent (SGD) optimizer, which was implemented in Pytorch.
The SGD optimizer had a learning rate of $0.03$ and a momentum of $0.9$ as parameters.
For model encryption, pixel shuffling with a block size of $M = 16$ was used as well as the patch size of ViT. The robustness of models was evaluated by using AutoAttack (AA) \cite{croce2020reliable} which consists of three white-box attacks (APGD-ce, APGD-t, FAB-t) and one black-box attack (Square).
The maximum perturbations for AA were ${\ell}_{\infty}$ and $\epsilon=8/255$ on the CIFAR-10 dataset.

\subsection{Experimental Result}
To confirm the effectiveness of the random ensemble, it was compared with a simple ensemble model (without random selection) where both ensembles consisted of four encrypted sub-models ($N=4$). \par
Table \ref{tab:CIFAR_1} shows experimental results where the clean accuracy indicates the accuracy of using clean test images without any adversarial noise.
From the table, Baseline (the use of plain models) was verified to be vulnerable to AEs.
The simple ensemble with encrypted sub-models was robust against the white-box attacks, but it was not robust against Square (black-box attack).
As a result, AutoAttack could easily mislead the model.  In contrast, the use of the random ensemble allows us to maintain high accuracy even when AutoAttack was applied.

\begin{table*}
    \centering
    \caption{Robust Accuracy on CIFAR-10 against AA (${\ell}_{\infty}, \epsilon=8/255, N=4$).}
    \begin{tabular}{c|c|ccccc}
    \hline
    \multirow{2}{*}{Model} & \multirow{2}{*}{Clean(\%)} & \multicolumn{5}{c}{Attack Method} \\
    & & APGD-ce(\%) & APGD-t(\%) & FAB-t(\%) & Square(\%) & AA(\%) \\
    \hline
    \hline
    ViT & \multirow{2}{*}{99.03} & \multirow{2}{*}{0.0} & \multirow{2}{*}{0.0} & \multirow{2}{*}{0.03} & \multirow{2}{*}{0.85} & \multirow{2}{*}{0.0} \\
    (baseline)& & & & & & \\
    \hline
    Simple & \multirow{2}{*}{98.27} & \multirow{2}{*}{98.34} & \multirow{2}{*}{98.18} & \multirow{2}{*}{98.27} & \multirow{2}{*}{1.54} & \multirow{2}{*}{1.54} \\
    Ensemble & & & & & & \\
    \hline
    Random & \multirow{2}{*}{98.20} & \multirow{2}{*}{98.22} & \multirow{2}{*}{98.23} & \multirow{2}{*}{98.33} & \multirow{2}{*}{\textbf{76.50}} & \multirow{2}{*}{\textbf{76.29}} \\
    Ensemble & & & & & & \\
    \hline
    \end{tabular}
    \label{tab:CIFAR_1}
\end{table*}

\vspace{-3mm}

\section{Conclusions}
We proposed a random ensemble of encrypted ViTs to improve the robustness against adversarial examples.
In the experiment, we used a benchmark attack, AutoAttack, to verify the effectiveness of the proposed method.
We confirmed that the proposed method is robust not only against white-box attacks but also against a black-box attack.

\vspace{-3mm}

\section*{Acknowledgment}
This study was partially supported by JSPS KAKENHI (Grant Number JP21H01327).

\bibliographystyle{IEEEtran}
\bibliography{main}

\begin{thebibliography}{10}
\providecommand{\url}[1]{#1}
\csname url@samestyle\endcsname
\providecommand{\newblock}{\relax}
\providecommand{\bibinfo}[2]{#2}
\providecommand{\BIBentrySTDinterwordspacing}{\spaceskip=0pt\relax}
\providecommand{\BIBentryALTinterwordstretchfactor}{4}
\providecommand{\BIBentryALTinterwordspacing}{\spaceskip=\fontdimen2\font plus
\BIBentryALTinterwordstretchfactor\fontdimen3\font minus \fontdimen4\font\relax}
\providecommand{\BIBforeignlanguage}[2]{{%
\expandafter\ifx\csname l@#1\endcsname\relax
\typeout{** WARNING: IEEEtran.bst: No hyphenation pattern has been}%
\typeout{** loaded for the language `#1'. Using the pattern for}%
\typeout{** the default language instead.}%
\else
\language=\csname l@#1\endcsname
\fi
#2}}
\providecommand{\BIBdecl}{\relax}
\BIBdecl

\bibitem{hitoshi2022an}
\BIBentryALTinterwordspacing
H.~Kiya, A.~P.~M. Maung, Y.~Kinoshita, S.~Imaizumi, and S.~Shiota, ``An overview of compressible and learnable image transformation with secret key and its applications,'' \emph{APSIPA Transactions on Signal and Information Processing}, vol.~11, no.~1, 2022. [Online]. Available: \url{http://dx.doi.org/10.1561/116.00000048}
\BIBentrySTDinterwordspacing

\bibitem{goodfellow2015explaining}
\BIBentryALTinterwordspacing
I.~J. Goodfellow, J.~Shlens, and C.~Szegedy, ``Explaining and harnessing adversarial examples,'' in \emph{3rd International Conference on Learning Representations, {ICLR} 2015, San Diego, CA, USA, May 7-9, 2015, Conference Track Proceedings}, Y.~Bengio and Y.~LeCun, Eds., 2015. [Online]. Available: \url{http://arxiv.org/abs/1412.6572}
\BIBentrySTDinterwordspacing

\bibitem{kurakin2017adversarial}
A.~Kurakin, I.~J. Goodfellow, and S.~Bengio, ``Adversarial machine learning at scale,'' in \emph{International Conference on Learning Representations}, 2017.

\bibitem{Hongyang2019theoratically}
H.~Zhang, Y.~Yu, J.~Jiao, E.~P. Xing, L.~E. Ghaoui, and M.~I. Jordan, ``Theoretically principled trade-off between robustness and accuracy,'' in \emph{Proceedings of the 36th International Conference on Machine Learning, {ICML} 2019, 9-15 June 2019, Long Beach, California, {USA}}, ser. Proceedings of Machine Learning Research, K.~Chaudhuri and R.~Salakhutdinov, Eds., vol.~97.\hskip 1em plus 0.5em minus 0.4em\relax {PMLR}, 2019, pp. 7472--7482.

\bibitem{maung2020encryption}
A.~MaungMaung and H.~Kiya, ``Encryption inspired adversarial defense for visual classification,'' in \emph{2020 IEEE International Conference on Image Processing (ICIP)}, 2020, pp. 1681--1685.

\bibitem{aprilpyone2021block}
------, ``Block-wise image transformation with secret key for adversarially robust defense,'' \emph{IEEE Transactions on Information Forensics and Security}, vol.~16, pp. 2709--2723, 2021.

\bibitem{tanaka2022on}
\BIBentryALTinterwordspacing
M.~Tanaka, I.~Echizen, and H.~Kiya, ``On the transferability of adversarial examples between encrypted models,'' in \emph{2022 International Symposium on Intelligent Signal Processing and Communication Systems (ISPACS)}, 2022, pp. 1--4. [Online]. Available: \url{https://doi.org/10.1109/ISPACS57703.2022.10082844}
\BIBentrySTDinterwordspacing

\bibitem{pang2019improving}
T.~Pang, K.~Xu, C.~Du, N.~Chen, and J.~Zhu, ``Improving adversarial robustness via promoting ensemble diversity,'' in \emph{Proceedings of the 36th International Conference on Machine Learning}, ser. Proceedings of Machine Learning Research, K.~Chaudhuri and R.~Salakhutdinov, Eds., vol.~97.\hskip 1em plus 0.5em minus 0.4em\relax PMLR, 09--15 Jun 2019, pp. 4970--4979.

\bibitem{yang2020dverge}
H.~Yang, J.~Zhang, H.~Dong, N.~Inkawhich, A.~Gardner, A.~Touchet, W.~Wilkes, H.~Berry, and H.~Li, ``Dverge: Diversifying vulnerabilities for enhanced robust generation of ensembles,'' in \emph{Proceedings of the 34th International Conference on Neural Information Processing Systems}, ser. NIPS'20.\hskip 1em plus 0.5em minus 0.4em\relax Red Hook, NY, USA: Curran Associates Inc., 2020.

\bibitem{maungmaung2021ensemble}
A.~MaungMaung and H.~Kiya, ``Ensemble of key-based models: Defense against black-box adversarial attacks,'' in \emph{2021 IEEE 10th Global Conference on Consumer Electronics (GCCE)}, 2021, pp. 95--98.

\bibitem{croce2020reliable}
F.~Croce and M.~Hein, ``Reliable evaluation of adversarial robustness with an ensemble of diverse parameter-free attacks,'' in \emph{Proceedings of the 37th International Conference on Machine Learning}, ser. ICML'20.\hskip 1em plus 0.5em minus 0.4em\relax JMLR.org, 2020.

\bibitem{madry2018towards}
A.~Madry, A.~Makelov, L.~Schmidt, D.~Tsipras, and A.~Vladu, ``Towards deep learning models resistant to adversarial attacks,'' in \emph{6th International Conference on Learning Representations, {ICLR} 2018, Vancouver, BC, Canada, April 30 - May 3, 2018, Conference Track Proceedings}.\hskip 1em plus 0.5em minus 0.4em\relax OpenReview.net, 2018.

\bibitem{carlini2017towards}
N.~Carlini and D.~Wagner, ``Towards evaluating the robustness of neural networks,'' in \emph{2017 IEEE Symposium on Security and Privacy (SP)}, 2017, pp. 39--57.

\bibitem{moosavi2016deepfool}
S.-M. Moosavi-Dezfooli, A.~Fawzi, and P.~Frossard, ``Deepfool: A simple and accurate method to fool deep neural networks,'' in \emph{2016 IEEE Conference on Computer Vision and Pattern Recognition (CVPR)}, 2016, pp. 2574--2582.

\bibitem{andriushchenko2020square}
M.~Andriushchenko, F.~Croce, N.~Flammarion, and M.~Hein, ``Square attack: A query-efficient black-box adversarial attack via random search,'' in \emph{Computer Vision -- ECCV 2020}, A.~Vedaldi, H.~Bischof, T.~Brox, and J.-M. Frahm, Eds.\hskip 1em plus 0.5em minus 0.4em\relax Cham: Springer International Publishing, 2020, pp. 484--501.

\bibitem{croce2020minimally}
F.~Croce and M.~Hein, ``Minimally distorted adversarial examples with a fast adaptive boundary attack,'' in \emph{Proceedings of the 37th International Conference on Machine Learning}, ser. ICML'20.\hskip 1em plus 0.5em minus 0.4em\relax JMLR.org, 2020.

\bibitem{croce2021robustbench}
F.~Croce, M.~Andriushchenko, V.~Sehwag, E.~Debenedetti, N.~Flammarion, M.~Chiang, P.~Mittal, and M.~Hein, ``Robustbench: a standardized adversarial robustness benchmark,'' in \emph{Thirty-fifth Conference on Neural Information Processing Systems Datasets and Benchmarks Track (Round 2)}, 2021.

\bibitem{mahmood2021robustness}
K.~Mahmood, R.~Mahmood, and M.~Van~Dijk, ``On the robustness of vision transformers to adversarial examples,'' in \emph{Proceedings of the IEEE/CVF International Conference on Computer Vision}, 2021, pp. 7838--7847.

\bibitem{krizhevsky2009learning}
\BIBentryALTinterwordspacing
A.~Krizhevsky and G.~Hinton, ``Learning multiple layers of features from tiny images,'' University of Toronto, Toronto, Ontario, Tech. Rep.~0, 2009. [Online]. Available: \url{https://www.cs.toronto.edu/~kriz/learning-features-2009-TR.pdf}
\BIBentrySTDinterwordspacing

\bibitem{dosovitskiy2021an}
A.~Dosovitskiy, L.~Beyer, A.~Kolesnikov, D.~Weissenborn, X.~Zhai, T.~Unterthiner, M.~Dehghani, M.~Minderer, G.~Heigold, S.~Gelly, J.~Uszkoreit, and N.~Houlsby, ``An image is worth 16x16 words: Transformers for image recognition at scale,'' in \emph{International Conference on Learning Representations}, 2021.

\end{thebibliography}
\end{document}